\documentclass[11pt, executivepaper]{article}
\usepackage[utf8]{inputenc}
\usepackage[T1]{fontenc}
\usepackage{natbib}
\usepackage{amsmath}
\usepackage{xcolor}
\usepackage{amsfonts}
\usepackage{graphicx}
\usepackage{enumitem}
\usepackage{geometry}
\usepackage{hyperref}
\hypersetup{colorlinks= true, allcolors=blue}
\setcitestyle{aysep={}}

\title{\textbf{Between Heisenberg and Feynman: Ettore Majorana's Reflections on the Meaning of Quantum Theory}}

\author{Andrea Oldofredi\thanks{Contact Information: University of Lisbon, Centre of Philosophy (Lancog Research Unit), Alameda da Universidade, 1600-214 Lisbon, Portugal. E-mail: aoldofredi@letras.ulisboa.pt}}

\begin{document}

\maketitle 

\begin{abstract}
Ettore Majorana is undoubtedly one of the most important figures in 20th-century Italian science. His contributions to nuclear physics and Quantum Mechanics (QM) included innovative ideas, and the search for Majorana-type neutrinos remains an ongoing experimental research question. However, very little is known about his perspective on the significance of the quantum state, the philosophical implications of the then-new mechanics, and the relationship between QM and other sciences. This chapter aims to address this gap. In more detail, here I present the standard interpretation of Majorana's view on the quantum state, according to which the Italian physicist anticipated the physical content of Feynman's path-integral formulation of quantum theory. In what follows, I critically examine this position and propose an alternative understanding of Majorana's beliefs on the nature of the quantum state and the wave function. Finally, I briefly discuss his reflections concerning the major conceptual revolutions introduced by QM, notably, in his view, the impossibility of extending the objectivity and determinism of classical mechanics to the quantum realm, and their consequences for the relation between physics and the social sciences, disciplines that Majorana deeply studied before his disappearance. 
\vspace{5mm}

{\em Accepted for publication in the volume ``Quantum Dilemmas: Philosophical Issues in Quantum Theory after One Century'', Arroyo, R.\ and Morganti, M.\ (eds), Springer}
\end{abstract}

\clearpage

\tableofcontents
\vspace{5mm}

\section{Introduction: A short biographical note}
\label{Sec1}

Ettore Majorana is undoubtedly one of the most important, iconic, and mysterious figures in 20th-century Italian science. His contributions to nuclear physics and (relativistic) Quantum Mechanics (QM) included revolutionary ideas\footnote{Such as the symmetric theory of electrons and positions, which led to the concept of Majorana fermions, i.e.\ those half-integer spin particles that are their own antiparticle.}, and the search for Majorana-type neutrinos remains an open area of experimental research. However, very little is known about his views on the meaning of quantum theory, especially regarding the significance of the wave function, the philosophical implications concerning objectivity and determinism, as well as the relationship between QM and other sciences. This chapter aims to address this gap. 
\vspace{2mm}

Given that Majorana's background may not be known to some readers, let us begin this chapter with an essential biographical note. Majorana was born in Catania, Italy, on August 5th, 1906, into a prestigious and wealthy family. After earning his diploma at the Liceo Classico ``Torquato Tasso'' in Rome in 1923, he studied engineering at the University of Rome until 1928. 
As the Nobel Laureate Emilio Segr\`e recalls, the several discussions between Majorana and him---who was studying engineering alongside Ettore---were crucial in convincing the latter to leave his Faculty and join the Royal Institute of Physics under Fermi's supervision, who immediately recognized the astonishing mathematical abilities of the new student (cf.\ \cite{Segre:1964}, see also \cite{Colangelo:2014}, pp.\ 43-44).\footnote{Segr\`e was the first to consider switching from engineering to physics. His meetings with Franco Rasetti and Enrico Fermi, appointed Professor of Theoretical Physics in 1926, greatly influenced his decision. Moreover, as underlined by Amaldi, between the fall of 1927 and the winter of 1928 Segr\`e informed Fermi about Majorana's extraordinary intellectual qualities \cite[p.\ XVII]{Bassani:2006}.} He eventually graduated in physics with the highest grades and honors, submitting the thesis \emph{The quantum mechanics of radioactive nuclei} in 1929.\footnote{A copy of Majorana's thesis is stored in his archival collection at Domus Galileana in Pisa, Italy.} 

Right after his graduation, Majorana joined Fermi's research group at the Institute of Physics in Rome. Between 1929 and 1933, he conducted theoretical research on several topics, including spectroscopy, particles with arbitrary spin, and thermodynamics. Many results obtained by Majorana in these years were published in academic journals\footnote{The English translation of all Majorana's published works, a commentary on them, and a detailed biographical note can be found in \cite{Bassani:2006}. This edited volume also includes his posthumous essay {\em The value of statistical laws in physics and social sciences}, which appeared in the academic journal {\em Scientia} only in 1942. As we will see, this article will be relevant for our investigation of Majorana's philosophical reflections on QM.}, whereas the vast amount of private notes he took in these years is archived at Domus Galileana in Pisa. As far as atomic physics is concerned, we should mention the essay {\em Oriented atoms in a variable magnetic field} (\cite{Majorana:1932b}), where Majorana ``evaluates the probability of spin-flip for atoms in a polarized beam in the presence of a varying magnetic field'' as underlined in \cite[p.\ 133]{Bassani:2006}. On the other hand, taking into account his work on quantum theory, the paper {\em Relativistic theory of particles with arbitrary intrinsic angular momentum} is particularly interesting (\cite{Majorana:1932b}). In his first work on elementary particles, Majorana attempted to generalize wave mechanics to relativistic regimes in a way that eliminates negative-energy solutions, which were instead notoriously present in Dirac's hole theory (\cite{Dirac:1930}, cf.\ for details \cite{Kragh:1990}, \cite{Oldofredi:2019b}). Moreover, in this essay, he derives a generalization of Dirac's equation able to describe relativistic particles with arbitrary spin.

Additionally, it is worth noting that Majorana's interest in nuclear physics increased in 1932 when the first ideas concerning the nucleus being composed solely of protons and neutrons were proposed (cf.\ e.g. \cite{Iwanenko:1932}).\footnote{Regarding this, Recami writes: ``As soon as news of Joliot-Curie's experiments reached Rome at the dawn of 1932, Ettore realized that they had discovered the ``neutral proton'' without realizing it. Even before the official announcement of the discovery of the neutron---made shortly after by Chadwick—Majorana was able to explain the structure and stability of atomic nuclei using protons and neutrons'' (\cite{Recami:2011}, p.\ 64, my translation from the original Italian).} By the end of March 1932, Amaldi recollects that Majorana developed a theory of light nuclei composed solely of protons and neutrons interacting through forces originating from the exchange in their spatial coordinates. However, he refused to publish it, as he believed his results were incomplete.\footnote{Amaldi admits that he and his colleagues are unsure whether Majorana discovered such an idea independently of Iwanenko's paper mentioned above.} Ettore outlined his theory at the Institute in Rome, and Fermi tried to convince him to publish his findings, even if partial, due to the inherent interest of such ideas. In the same year, in fact, Heisenberg published an influential essay on the atomic nucleus including ideas similar to Ettore's (\cite{Heisenberg:1932}). Although Fermi's efforts to persuade Majorana to publish his theory were unsuccessful, he was able to convince his former student to visit prestigious institutions in Europe. 
In 1933, Majorana visited two of the leading centers for theoretical physics at the time. At the end of January 1933, he joined Werner Heisenberg in Leipzig until March, then Niels Bohr's group in Copenhagen. Subsequently, the Italian scientist returned to Leipzig before going back to Rome in the fall of that year. Majorana's visit to Leipzig proved somewhat productive; he not only established a very good relationship with Heisenberg, but the latter was able (also in virtue of his authority) to motivate Majorana to publish his results on exchange forces, which included a correction of Heisenberg's own theory (\cite{Majorana:1933a, Majorana:1933b}).  

Due to health issues that began during his visit to Heisenberg, and that were probably related to nervous exhaustion (cf.\ Amaldi in \cite[p.\ XL]{Bassani:2006}), Majorana started to come to the Institute in Rome less frequently; he also did not go there for entire months. From 1934 to 1937, indeed, Majorana spent a lot of time at his house, where he deeply studied philosophy (Schopenhauer's works in particular), social sciences, and medicine besides mathematics and physics. Pushed by Fermi and his friends at the Institute of Physics, Majorana decided to enter the competition for the opening chair in theoretical physics at the University of Palermo in 1937. Given that Majorana had not published any new papers since 1933, he was advised to publish new material to increase the probability of success---at that time, in fact, Majorana's scientific production was rather meager. 

Referring to this, in 1937, he published his most celebrated and well-known article {\em A symmetric theory of electrons and positrons} that appeared in the journal {\em Il Nuovo Cimento}, although it seems that the first ideas and results of this article were actually conceived and obtained between 1932 and 1933, when the positron was experimentally discovered (\cite{Majorana:1937}). 
In this work, Majorana reformulated Dirac's theory, eliminating negative-energy states, and providing a new field-theoretic description of electrons and positrons that led to the notion of Majorana fermions, i.e., those half-integer spin particles that are their own antiparticle, as already stated in footnote 1. Importantly, Majorana suggests that neutrinos can be particles of this type, opening a new experimental research area that tries to find such ``Majorana-type'' neutrinos.

The board of examiners for the competition for the Chair in theoretical physics in Palermo, which included Fermi, met in October 1937. The committee, at the request of Senator Giovanni Gentile and in agreement with him, decided to write a letter to the Minister of National Education, Giuseppe Bottai, to approve a direct nomination of Majorana as Professor of Theoretical Physics at the Royal University of Naples. In a few weeks Bottai accepted, and Majorana was officially appointed on November 16, 1937.\footnote{Such a request was based on Article 8 of the Royal Decree-Law, 20 June 1935, n.\ 1071, used to appoint Guglielmo Marconi as Professor at the University of Rome. This law allowed the direct nomination of a candidate based on their scientific excellence. For details cf.\ \cite{Bassani:2006}, pp. XXIV-XXV and \cite{Esposito:2006}.} 

In January 1938, Ettore moved to Naples, where he delivered an inaugural lecture for the academic staff of the Institute of Physics and (more generally) for the staff of the University on January 13th. This marked the beginning of his course in Theoretical Physics, which covered several topics such as old quantum theory and its application to atomic physics\footnote{As Esposito correctly underlines, this part of the course is heavily influenced by Fermi's book {\em Introduzione alla fisica atomica} (Introduction to atomic physics), used in the courses attended by Majorana in Rome (\cite{Fermi:1928}).}, special relativity, and the new formulation of QM together with the necessary mathematical machinery to properly understand it, e.g.\ matrix theory, vector spaces, and operator theory. Moreover, according to \cite{Esposito:2006b, Esposito:2006c}, upon request of Luciano Carrelli, then director of the Institute of Physics, in February 1938 Majorana allegedly delivered a conference on advanced topics in quantum theory, particularly the quantum description of molecular bonding. However, all his activities ended prematurely and unexpectedly at the end of March 1938, precisely on Thursday, the 24th, when Majorana mysteriously disappeared.\footnote{The last certain information about him is dated 26 March 1938, when he sent a letter to Carrelli. As soon as the news of his disappearance circulated, his family, as well as his colleagues in Naples and Rome, started tirelessly to find him. As is well known, unfortunately, these efforts were in vain. Many speculations and hypotheses concerning Majorana's disappearance abound; a careful analysis of this matter is given in \cite{Sciascia:1975}. For a detailed biography of Majorana and a review of his scientific achievements, see \cite{Recami:2011}. The latter volume also contains copies of letters, documents, and recollections from Majorana's colleagues.}

As the reader will observe, the preparatory notes of Majorana's last conference\footnote{A detailed archival history of the discovery of Majorana's preparatory notes for his conference is provided in the introduction of \cite{Esposito:2006}. An English translation of this text, together with a detailed historical introduction, is included in \cite{Esposito:2006b, Esposito:2006c}.} and his lectures on theoretical physics at the University of Naples---among other documents---will be crucial for our analysis for two main reasons. On the one hand, in this conference he discussed the meaning of the quantum state, a topic rarely addressed elsewhere. On the other hand, Majorana's last two lectures focused on the modern version of quantum mechanics, during which he also provided interesting comments about the interpretation of the wave function, the central object of the theory.

Against this background, the present essay is structured as follows.  I introduce and critically examine the claims about Majorana's perspective on the meaning of the quantum state found in \cite{Esposito:2006b, Esposito:2006c}, respectively, in Section \ref{Sec2} and Section \ref{Sec3}. Specifically, I question the conjecture according to which Majorana would have anticipated the physical content of Feynman's path integral formulation of QM. In Section \ref{Sec4}, I propose a new hypothesis concerning the interpretation of the notions of quantum state and wave function that aligns more closely with Majorana's texts. To support my arguments, I will employ his lecture notes on theoretical physics, a document I believe is essential for understanding Majorana's perspective correctly, but has regrettably been overlooked in historico-philosophical discussions of his work. Still in Section \ref{Sec4}, I will review and discuss Majorana's perspective on the philosophical implications of QM and the relationship between this framework and other branches of science, particularly the social sciences. Finally, Section \ref{Sec5} contains conclusive comments.

\section{Anticipating Feynman: The Path-Integral Conjecture}
\label{Sec2}

Examining the historical literature on Majorana and his contributions to theoretical physics, it is very common to find claims that he was a forerunner, with physical and mathematical intuitions ahead of his time, anticipating ideas that would only be discovered years later, or that his work---and its significance---would only be fully recognized and understood several years after its publication.\footnote{Various examples can be found in \cite{Recami:2011}, in Amaldi's recollections of Majorana's life and works in \cite{Bassani:2006} as well as in \cite{Amaldi:1966}.} In the following, we will introduce another example that can be ascribed to this general pattern. 

Since one of our goals in this essay is to understand Majorana's perspective on the concepts of quantum state and the meaning of the wave function, let's explore this question without any further ado. The natural place to begin my analysis is the preparatory text for the conference, or {\em lectio magistralis}, that Majorana likely delivered in winter 1938 (between the end of January and early February) at the University of Naples, because, as mentioned in the previous section, he explicitly planned to discuss the notion of quantum state on that occasion. Indeed, the historical importance of this document lies in the fact that Majorana included a few comments about the interpretation of the then-new theory of mechanics and its relationship with the classical framework.

According to Esposito\footnote{Esposito credited the Italian physicist Nicola Cabibbo as the one who first saw a resemblance to Feynman's ideas in Majorana's text.}, who first edited, commented, and translated the text of Majorana's conference, such notes would contain an anticipation of the physical content of Feynman's path-integral interpretation of QM (\cite{Feynman:1948}). More precisely, Majorana's interpretation of the quantum state, in Esposito's opinion, provides the foundations of Feynman's celebrated approach. This emerges clearly from the following quotes:

\begin{quote}
[I]t is without doubt that no trace can be found of the formalism underlying the Feynman path-integral approach to quantum mechanics in the Majorana manuscript [...]. Nevertheless, \emph{it is very interesting that the main physical items, about the novel way of interpreting the theory of quanta, were realized well in advance by Majorana} (\cite{Esposito:2006b}, p.\ 1149, my emphasis).
\vspace{3mm}

Coming back to the Majorana paper here considered, we immediately realize that it contains nothing of the mathematical aspect of the peculiar approach to Quantum Mechanics [of Feynman]. Nevertheless, an accurate reading reveals, as well, the presence of the \emph{physical} foundations of it (\cite{Esposito:2006c}, p.\ 212).
 \end{quote}

\noindent The textual evidence that Esposito employs to support his interpretation is taken from the first section of the preparatory notes. Discussing the notion of quantum state, Majorana claims that

\begin{quote}
[a]ccording to the Heisenberg theory, a quantum state corresponds not to a strangely privileged solution of the classical equations but rather to a set of solutions which differ for the initial conditions and even for the energy, what is meant as precisely defined energy for the quantum state corresponds to a sort of average over the infinite classical orbits belonging to that state. Thus the quantum states come to be the minimal statistical sets of classical motions, slightly different from each other, accessible to the observations. These minimal statistical sets cannot be further partitioned due to the uncertainty principle, introduced by Heisenberg himself, which forbids the precise simultaneous measurement of the position and the velocity of a particle, that is the determination of its orbit (Majorana quoted in \cite{Esposito:2006c}, p.\ 216).
\end{quote}

Esposito's intuition is grounded in the statement that the quantum state corresponds to a set of solutions differing for initial conditions and energy, where ``what is meant as precisely defined energy for the quantum state corresponds to a sort of average over the infinite classical orbits belonging to that state''. Taking into consideration the path-integral formulation of QM, the reasons for which Esposito finds an anticipation of Feynman's ideas in the passage quoted above are immediately evident. Indeed, the path-integral approach can be essentially described by the following formula
\vspace{2mm}

\begin{equation}
\langle x_f, t_f|x_i, t_i\rangle=\int\mathcal{D}x(t)e^{i\mathcal{S}[x(t)]/\hbar}.
\end{equation}
\vspace{2mm}

\noindent On the l.h.s.\ of this equation, $\langle x_f, t_t|x_i, t_i\rangle$ represents the transition amplitude of a quantum particle ``traveling'' in space from an arbitrary initial point $x_i$ at time $t_i$ to a final point $x_f$ at time $t_f>t_i$---this is in fact called {\em propagator}. On the r.h.s., $\mathcal{D}x(t)$ represents the integration over all possible paths $x(t)$ of the particle between $x_i$ at time $t_i$ and $x_f$ at time $t_f$. Finally, $e^{i\mathcal{S}[x(t)]/\hbar}$ provides the weights of all such possible paths, where $S$ is the classical action. Notably, according to Feynman, the probability amplitude on the l.h.s.\ is defined taking into account all the contributions of all paths in configuration space, where each contribution is proportional to the quantity $e^{i\mathcal{S}[x(t)]/\hbar}$.
\vspace{3mm}

  \begin{center}
    \includegraphics[scale=0.8]{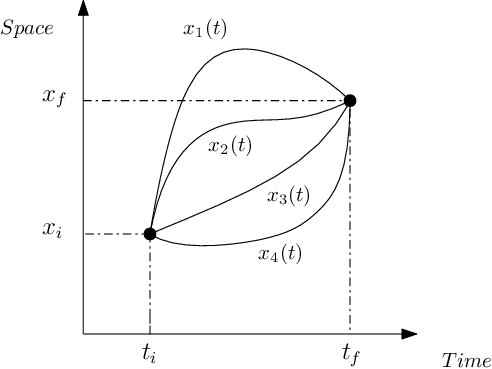}
    \begin{quote}
\footnotesize{Idealized representation of a few possible paths that a quantum particle can follow from point $x_i$ to $x_f$.}
\end{quote}
  \end{center}
\vspace{3mm}

\noindent Against this background, Esposito affirms that Majorana's phrase ``solutions which differ for the initial conditions'' should be interpreted in Feynman's terms as referring to different integration paths:

\begin{quote}
 Let us note that the ``solutions which differ for the initial conditions'' correspond, in the Feynman language of 1948, precisely to the different integration paths. In fact, the different initial conditions are, in any case, always referred to the same initial time ($t_i$), while the determined quantum state corresponds to a fixed end time ($t_f$) (\cite{Esposito:2006c}, pp.\ 212-213).  
 \end{quote}

\noindent The author moreover states that the expression ``slightly different classical motions'' corresponds to ``the sufficiently wide integration region for quantum systems''.\footnote{Esposito claims: ``The introduced issue of ``slightly different classical motions'' [...], is thus evidently related to that of the sufficiently wide integration region for quantum systems'' ({\em op.cit.}).}

Finally, the last piece of textual evidence used to argue for the anticipation of Feynman's ideas is taken from the third and last part of Majorana's note, where he discusses the theory of homopolar valence according to the method of bonding electrons, by studying the motion of an electron in the field of two hydrogen nuclei. Towards the end of the section, Majorana speaks about the wave function of an electron, and says that it ``corresponds in Quantum Mechanics to any possible state of the electron'' (Majorana in \cite{Esposito:2006c}, p.\ 222). Referring to this specific sentence, Esposito claims that one should not interpret it in the usual statistical interpretation of the quantum mechanical wave function, according to which $\psi$ contains the maximal information available on the quantum system it represents. Rather, it should be read ``in the meaning given by Feynman'', i.e., in terms of path integrals.\footnote{In fact, we read: ``And, in a later passage, it is more explicitly stated that the wave function ``corresponds in Quantum Mechanics to any possible state of the electron''. Such a reference, that only superficially could be interpreted, in the common acceptation, that all the information on the physical systems is contained in the wave function, {\em should instead be considered in the meaning given by Feynman}'' (\cite{Esposito:2006c}, p.\ 213, my emphasis).} 

\section{A critical discussion of Esposito's interpretation}
\label{Sec3}
 
Although the interpretation of Majorana's text introduced earlier is certainly intriguing---and, to some extent, audacious---in the following, I argue that such an understanding is questionable. To start our discussion, let's critically examine the claim that Majorana anticipated the physical foundations of Feynman's path-integral approach to QM. As we will see, another interpretation of this document is possible, one in which no anticipation of Feynman's ideas is found.

To accomplish this, let's analyze the textual evidence used by Esposito in more detail, which is drawn from the first section of the notes where Majorana discusses the concept of quantum state. It is important to note that to understand the passage quoted by Esposito correctly, one should place it within Majorana's broader discussion on the relationship between classical and quantum descriptions of physical systems. 

The document starts indeed with a simple remark, namely that in QM the internal energy of a closed system (e.g., atoms or molecules) can take discrete eigenvalues $E_0, E_1, E_2, \dots$ and that to each value is associated a quantum state in which the system can ``remain indefinitely without external perturbations''. Then, Majorana asks what concept would correspond in classical mechanics to the notion of quantum state, given that in this section the author emphasizes the new features of QM as well as the limits of a classical treatment of microscopic systems. He then answers this question by pointing out that in the classical theory, a system composed of $N$ point particles is completely described when their coordinates $q_1, \dots, q_n$ are known as a function of time $q_i=q_i(t)$. The latter equation, however, is no longer valid for subatomic, atomic, and molecular systems, thereby modifications of classical mechanics have been proposed. On the one hand, Majorana mentions Bohr-Sommerfeld's old quantum theory; on the other hand, the then-emerging novel quantum mechanics, which is the framework on which he focuses in the rest of his text. It is moreover important to underline that in Majorana's opinion, not only is QM ``primarily due to Heisenberg'', but also that his approach is ``substantially closer to the classical conceptions than the old one''---i.e., the theory of Bohr and Sommerfeld---that is seen as an {\em ad hoc}, hybrid proposal in which quantization rules are superimposed on classical trajectories. 

The first sentence of Majorana's quote tells us, in fact, that
\begin{quote}
According to the Heisenberg theory, a quantum state corresponds not to a strangely privileged solution of the classical equations...
\end{quote}

\noindent As is well-known, conforming to the old quantum theory, classical mechanics is still valid ``in the atom'', e.g., the electron's motion around the nucleus of the Hydrogen atom is still described via a solution of classical dynamical equations. However, not all classical solutions are realized in nature, but only a discrete infinity of them satisfying the quantization conditions. Here, Majorana underlines that, contrary to the framework of Bohr and Sommerfeld, a quantum state does not correspond to particular classical solutions admitted by the quantization rules. Rather, it amounts to 

\begin{quote}
...\ a set of solutions which differ for the initial conditions and even for the energy, what is meant as precisely defined energy for the quantum state corresponds to a sort of average over the infinite classical orbits belonging to that state. 
\end{quote}

\noindent This sentence explicitly mentions Heisenberg's theory, providing a clue about which form of quantum mechanics is being discussed---specifically, what is now called the Heisenberg picture of QM. This formulation is based on the work of the Göttingen school---in particular Max Born, Werner Heisenberg, and Pascual Jordan---in the mid-1920s. Understanding this detail is crucial to grasping the meaning of the quoted passage. 

The foundations of Heisenberg's theory lie in his pioneering work {\em Über quantentheoretische Umdeutung kinematischer und mechanischer
Beziehungen} (\cite{Heisenberg:1925}), where he aimed at providing a new quantum mechanical interpretation of classical kinematics and mechanics, as clearly stated in the paper's title. In order to do so, the German physicist begins his project with a methodological remark, namely that his theory will be based only on observable quantities as the energy of stationary states and their associated frequencies, whereas unobservable quantities, such as an electron's orbit or its position, are left out of the physical description (cf.\ \cite{Aitchison:2004} for details). 

To translate this methodological claim into practice, Heisenberg needs to provide a new interpretation of the concept of an electron's position, and of its dynamical motion. The first task is achieved as follows.\footnote{I am here following \cite{Fedak:2009} reconstruction of Heisenberg's reasoning.} 
In this work, electrons are no longer modeled as point particles moving along classical trajectories; they are instead described as a sum of virtual oscillators. Heisenberg takes into account one-dimensional periodic systems, whose classical motion is described by the time-dependent position $x(n, t)$ (where $n$ represents a stationary state), representing this periodic function in terms of a Fourier series:
\begin{align}
\label{Hei}
x(n, t)= \sum_{\alpha}a_{\alpha}(n)e^{i\alpha\omega(n)t}.
\end{align}

Now, the $\alpha$th Fourier component related to the stationary state $n$ has amplitude $a_\alpha(n)$ and frequency $\alpha\omega(n)$, which are in fact observable quantities---in particular, Heisenberg observed that frequencies depend on two terms. He then realized that, taking into consideration Bohr's correspondence principle, transition rates between quantum energy levels correspond to Fourier coefficients of classical electron orbits. More precisely, the $\alpha$th Fourier component related to the stationary state $n$ corresponds to the quantum jump from the state $n$ to the state $n-\alpha$. Thus, relying on this principle

\begin{quote}
Heisenberg replaced the classical component $a_{\alpha}(n)e^{i\alpha\omega(n)t}$ by the transition component $a(n,n-\alpha)e^{i\omega(n-n-\alpha)t}$. [...] Unlike the sum over
the classical components in Eq.\ \eqref{Hei} Heisenberg realized that a similar sum over the transition components is meaningless. Such a quantum Fourier series could not describe the
electron motion in one stationary state ($n$) because each term in the sum describes a transition process associated with two states ($n$ and $n-\alpha$) (\cite{Fedak:2009}, p.\ 136).
\end{quote}

\noindent Later in this work, Heisenberg eventually completed his first conceptual revolution and reinterpreted the notion of an electron's position in terms of the set of {\em transition components} 
\begin{align}
x\longrightarrow{a(n, n-\alpha)e^{i\omega(n-n-\alpha)t}.}
\end{align}

\noindent That is, Heisenberg's approach does not eliminate the notion of classical orbit, but {\em redefines} it by taking the terms of the Fourier series and organizing them into a table, where instead of a single state $n$, there is an item including a transition from $n$ to $n-\alpha$.

A second breakthrough of \cite{Heisenberg:1925} is the recognition that a collection of virtual oscillators becomes a sum of Fourier coefficients in the limit of high quantum numbers, and in that paper, Heisenberg discovered a method to multiply different collections of virtual oscillators, realizing that their product is not commutative. Born later understood that such collections are nothing but matrices---that were then unknown to the young Heisenberg.\footnote{In this paper and in subsequent essays with Born and Jordan, Heisenberg also provided a full quantum dynamics, but this topic lies outside the scope of the present chapter.}

In the successive works on matrix mechanics (\cite{Born:1925}, \cite{Born:1926}), then, the position of an electron is represented by an infinite matrix:

\begin{align}
X_{nm}= \begin{pmatrix}
x_{11} & x_{12} & x_{13} & \cdots & x_{1m}\\
x_{21} & x_{22} & x_{23} & \\ 
x_{31} & x_{32} & x_{33} & & \vdots\\
\cdots & & &  \ddots \\
x_{n1} & x_{n2} & x_{n3}& \cdots & x_{nm} \\
\end{pmatrix}.
\end{align}
\vspace{2mm}

\noindent In such a matrix, elements of each row and column correspond to energy states of a system, where the index represents initial and final states of quantum transitions from a given energy level to another---the index $n$ (row) indicates the final quantum state, whereas the index $m$ (column) expresses the initial quantum state. In more detail, the off-diagonal elements $X_{nm}$ correspond to transitions, to which is associated a certain probability amplitude that the electron jumps from a state $m$ to a state $n$. Such elements indeed specify the amplitude with which the radiation of a certain frequency is emitted; since radiation is emitted in discrete quanta when electrons transition from one energy level to another, the oscillator amplitude denotes the rate at which such a transition takes place.\footnote{As is well-known, the frequency of light emitted in such a quantum jump obeys Bohr's condition \begin{align}
\label{bohr}
\nu_{nm}= \frac{E_n-E_m}{h}.
\end{align}} Moreover, the quantity $|x_{nm}|^2$ is proportional to the probability that such a quantum jump takes place. On the other hand, the diagonal elements correspond to stationary states, where quantum jumps do not occur.

In Heisenberg's theory, thus, the position of an electron cannot be given by a single scalar quantity; rather, the position matrix describes all possible states (with relative transitions) in which the electron may be. Thereby, the notion of orbit is translated into a matrix expressing all the possible combinations of quantum jumps.

Referring to this, Heisenberg realized that a state $n$ is defined by each row of a position matrix, where the matrix elements correspond to all the transitions from one initial energy level. The quantum state is, therefore, a {\em synthesis} of all possible transitions that start from each energy level. As we already underlined above, from the correspondence principle it follows that the higher-order transitions from a certain orbit correspond (in the classical limit) to higher harmonics of that orbit. Thus, each row can be interpreted as all the Fourier components of a single orbit.

Now, Majorana's claim, according to which energy is the ``average over the infinite classical orbits belonging to that state'' can be traced down to the fact that, in the limit of high quantum numbers, the frequency $\nu_{nm}$ tends to the classical frequency (cf.\ footnote 16). However, whereas in classical physics energy is defined by the initial position and momentum of a given system, in Heisenberg's theory, the energy of a state $n$ is the result of a weighted sum of terms (i.e., the elements of the matrix $X_{nm}$). From a formal perspective, this fact ``averages'' those orbits that would be classically similar to each other, to obtain a unique discrete value $E_n$ for the energy. 

To see this fact more clearly, let us consider a simple example given by Majorana a few lines after the passage quoted by Esposito. He considers a harmonic oscillator with frequency $\nu$ that can oscillate classically with arbitrary amplitude and phase; its energy is given by $E= 2\pi^2m\nu^2A^2_0$, where $m$ is the mass and $A_0$ the maximal elongation. According to QM, possible values of $E$ are $E_0=1/2h\nu$, $E=3/2h\nu$, ... $E_n=(n+1/2)h\nu$, etc. The crucial remark made by Majorana is that
 
\begin{quote}
in this case we can say that the fundamental state with energy $E_0=1/2h\nu$ roughly corresponds to {\em all classical oscillations with energy comprised between} 0 and $h\nu$, the first excited state $E=3/2h\nu$ corresponds to the classical solutions with energy comprised between $h\nu$ and $2\cdot h\nu$ and so on (Majorana in \cite{Esposito:2006c}, p.\ 129, my emphasis).
\end{quote}

\noindent In this example, Majorana claims that each oscillation can only be connected to one energy state. It is, in fact, the energy of the oscillation itself that determines whether it should be grouped into each energy state. Therefore, the space of all possible oscillations is partitioned into {\em disjoint sets}, each corresponding to a single energy state. That is, if we were to be given a classical oscillation, we would know which quantum energy state it would correspond to. When one computes path integrals, however, one has to compute them on all possible paths for all possible states. Conversely, every possible path must be included for any quantum state. Therefore, this idea of partitioning, which seems clearly expressed by Majorana, is not present in the concept of the path integral. As a consequence, Esposito's interpretation seems to be technically inadequate.\footnote{I am grateful to Gabriele Carcassi for pointing out this technical fact in private communication.}

Let me now summarize our discussion so far. According to Heisenberg's picture, the quantum state does not describe a single classical orbit, but, as exemplified by Majorana in the above quotation, it refers to a set of orbits. The latter, in turn, should be translated into the jargon and concepts of matrix mechanics, given that---to be conceptually rigorous---in Heisenberg's theory the very notion of classical trajectory loses its physical meaning, as already stated at the beginning of our discussion. Moreover, the energy of a quantum state is given by an average computed by taking into account an infinite number of classical orbits that slightly differ in their initial conditions (positions and momentum). Now, such differences are not further analyzable in virtue of the inherent constraints posed by quantum theory, in particular by Heisenberg's position-momentum uncertainty relation.

Referring to this, let us analyse the last part of Majorana's passage, which says that
\begin{quote}
... the quantum states come to be the minimal statistical sets of classical motions, slightly different from each other, accessible to the observations. These minimal statistical sets cannot be further partitioned due to the uncertainty principle, introduced by Heisenberg himself, which forbids the precise simultaneous measurement of the position and the velocity of a particle, that is the determination of its orbit.
\end{quote}

\noindent In order to understand this statement, let us mention that, contrary to the classical case where a trajectory is uniquely determined once position and momentum are known, quantum theory poses inherent limitations to the simultaneous measurements of such quantities.\footnote{It is indeed well-known that the notion of trajectory loses its meaning in QM.} Majorana here hints at a very well-known consequence of Heisenberg uncertainty relations. Namely, to the fact that, contrary to the classical case where a physical system can be described with infinite precision as a point in phase space, in quantum theory, there exists an inherent limitation to what can be known about a certain quantum object; therefore, the quantum state is the minimal volume that a quantum particle can occupy in phase space. However, such volume can never be a single point (cf.\ figure below).

  \begin{center}
    \includegraphics[scale=0.6]{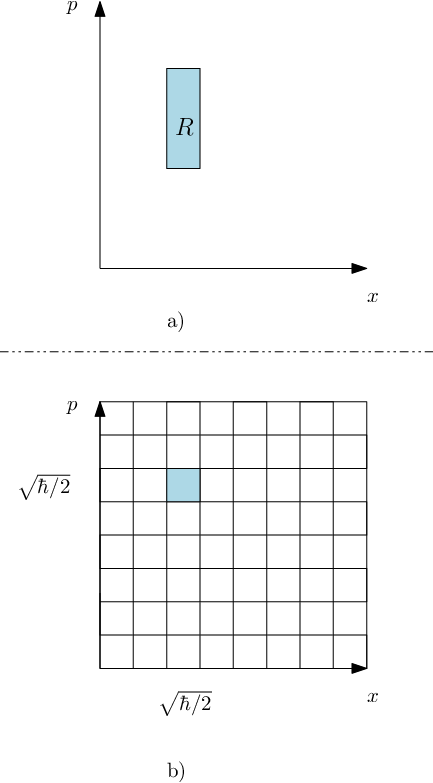}
    \begin{quote}
\footnotesize{In QM, position and momentum cannot be measured simultaneously with infinite precision; thus, the quantum state cannot be a point in phase space. Figure a) shows a region occupied in phase space by a quantum system, whereas Figure b) displays the smallest cells of such a phase space.}
\end{quote}
  \end{center}
\vspace{3mm}

Referring to this, in his lecture notes on theoretical physics---precisely in his 21st lecture---Majorana underlines very clearly the limitation of a classical description of microscopic systems:

\begin{quote}
  \vspace{2mm}
The uncertainty relations, however, mark not only the limits placed on the possibility of simultaneous observation of $p$ and $q$, but also the limit of the classical mechanics of the free point (principle of inertia). Indeed, it is not possible to interpret the motion of the probability $\bar{\psi}\psi$ according to classical conceptions, assuming that it is possible to associate to each $\psi$ a function $S(p,q) \geq 0$ representing the probability that the electron has an initial position and velocity and calculating the probabilities of future positions via the principle of inertia. Only if the wave function is largely undetermined and consequently instead of the uncertainty relations the following holds $$ \Delta x \Delta p \gg h$$ one can legitimately use classical mechanics, i.e.\ it is possible to define with sufficient approximation the classical statistical ensemble $S(p, q)$ and make use of it for the future prediction of the material point applying the principle of inertia (\cite{Esposito:2006}, p.\ 124, my translation from the original Italian).
  \end{quote}
  \vspace{2mm}

Concluding, in light of our brief discussion about the fundamentals of Heisenberg's theory, it seems plausible to affirm that Majorana explained compactly and elegantly what the notion of quantum state is according to Heisenberg's approach to QM, and what relation actually exists between a quantum and classical description of physical systems. If the interpretation offered in this section is correct, then it would be historically (and technically) inaccurate to claim that Majorana anticipated Feynman's path-integral formulation of quantum theory. 

\section{The meaning of $\psi$ and the implications of Quantum Mechanics}
\label{Sec4}

Having examined the shortcomings of Esposito's interpretation and proposed an alternative understanding of Majorana's quote studied earlier, let's examine in more detail his view on the meaning of the wave function and the conceptual innovations introduced by QM. 

To proceed with our discussion, it is useful to consider the last point raised by Esposito at the end of Section \ref{Sec2}, namely that the notion of wave function should be understood in Feynman's terms (cf.\ footnote 13). The problem studied by Majorana in the final section of this preparatory note concerns the description of an electron in a field of two hydrogen nuclei. More precisely, he discusses at length the difference in the treatment of such an electron between classical and quantum mechanics, and emphasizes that resonance forces present in QM, which are essential to describe chemical bonding, have no analogue in the classical theory---which is indeed inadequate for the task. For an exact computation of the electron's behavior, he underlines, we must use quantum theory and introduce the notion of wave function, which, as Majorana says, corresponds to ``any possible state of the electron''. Contrary to Esposito's thesis, however, $\psi$'s significance must be intended according to the usual Born interpretation, for Majorana explicitly claims that its squared modulus gives the probability that a certain quantum object---in this case, the electron in the H$_2$ molecule---``lies in the volume unit around a generic point $x$, $y$, $z$'' (Majorana in \cite{Esposito:2006c}, p.\ 223). Majorana then continues by saying that the quantum wave function must obey Schr\"odiger equation and ``thus we can always multiply $\psi$ in any point by a fixed real or complex number of modulus 1, this constraint being required by the normalization condition $\int |\psi|^2$d$x$d$y$d$z=1$ which is necessary for the mentioned physical interpretation of $|\psi|^2$'' (Majorana in {\em op.\ cit.}). It is therefore evident that the Italian physicist is referring to the statistical view of the wave function; consequently, there is no room for hermeneutic debates or interpretational issues regarding the significance of $\psi$ in this specific passage. 

Majorana's lecture notes provide further evidence of the last claim. After introducing the students to Einstein's special theory of relativity and to Compton's effect, in March 1938 he lectured on the then-new formulation of quantum mechanics, speaking about the Schr\"odinger picture, rather than Heisenberg's formulation mentioned in the previous section. Notably, before a proper treatment of the new theory, a few classes were focused on the relevant mathematical background, where elements of matrix calculus, vector spaces, operator theory, and Fourier analysis were introduced. Subsequently, Majorana outlined de Broglie's wave mechanics in detail, and this framework was used as the foundation to illustrate Schr\"odinger's approach to QM. Indeed, in his twentieth lecture---on Tuesday, March 22nd, 1938---Majorana explained Schr\"odinger's wave equation, focusing on non-relativistic quantum theory in which one can represent a material point, as e.g.\ an electron, using the notion of wave function $\psi$. In this lecture, Majorana claims that according to Schr\"odinger himself, the corpuscular representation of the electron must be abandoned in favor of an undulatory description. According to the Austrian physicist, an electron would be defined as a diffuse electric charge with density $\rho=-e\bar{\psi}\psi$.\footnote{This is the formalism employed by Majorana. The bar over $\psi$ represents complex conjugation; in modern notation, it corresponds to $\psi^*$.} However, Majorana explains in detail why such an interpretation of the electron is physically incorrect\footnote{To this regard, Majorana says: ``This interpretation, however, runs into difficulties similar to those encountered when one tries to represent a quantum of light by means of an electromagnetic vibration extended to a limited region of space, since a wave packet tends to dissolve over time by increasing its dimensions and can also be quickly divided in two by making it collide against the edge formed by two reflecting surfaces, while experience (Millikan) proves that just as the energy $h\nu$ of a quantum of light so also the charge $-e$ of the electron is an indivisible unit'' (Majorana in \cite{Esposito:2006}, p.\ 118, my translation from the original Italian).}, and claims:

\begin{quote}
For this reason [the physical inadequacy of Schro\"odinger ideas], M.\ Born (1927) proposed the statistical interpretation of the wave function $\psi$, according to which the electron is actually a point-like corpuscle and $\bar{\psi}\psi$ represents not the density of a diffuse electric charge, but the probability that the electron is located at a certain point. The function $\psi$ does not consequently have the objective classical meaning of a diffuse vibration but it is only a mathematical tool to summarize our knowledge on the state of the electron which can be deduced from previous experiences.
[...]
 We will see how $\psi$ can be deduced from previous experiences, and that there are arguments for supposing that the probabilistic link (nexus) established between past and future experiences represents the maximal possible extent of our knowledge of nature (\cite{Esposito:2006}, p.\ 118, my translation from the original Italian).
  \end{quote}

The first thing to note is that, once again, Born's statistical interpretation is mentioned as the correct reading of $\psi$. Second, from this quote, we see that the notion of wave function is interpreted as a mathematical device that contains the maximal knowledge we have of a certain quantum object, and that is useful to compute probabilities on future measurements performed on it. Since Majorana affirms that the probabilistic knowledge given by the wave function is the maximal information we can have on microscopic components of reality, it is plausible to claim that he had an {\em epistemic} conception of $\psi$. Thus, the wave function is an abstract, formal notion without a direct physical meaning. 
In addition, it is relevant to say that whereas the wave function pertains to epistemology, the notion of quantum state, or in this case the state of the electron, is {\em ontic}, for it refers instead to the actual physical state of affairs of a certain system. The squared modulus of $\psi$ provides, in fact, the probability that a given object is {\em actually located} somewhere in space and time. Therefore, the notions of quantum state and wave function are not equivalent. In sum, we can safely conclude that Majorana did not anticipate Feynman's path integral formulation of quantum theory, but rather held an epistemic interpretation of the wave function.

Referring to the ontological content of QM instead, in the posthumous essay {\em The value of statistical laws in physics and social sciences} (cf.\ footnote 4 for bibliographical details), Majorana claimed that QM entails a novel conception of matter, and this is one of the two major innovations introduced by this theory.
The Italian scientist claims that in quantum theory, we find

\begin{quote}
[a] certain lack of objectivity in the description of phenomena. Any experiment carried out on an atomic system exerts a finite disturbance on it that cannot, for reasons of principle, be eliminated or reduced. The result of any measurement, therefore, seems to concern the state to which the system is brought during the experiment itself rather than the unknowable state in which it was before being disturbed. This aspect of quantum mechanics is undoubtedly more disturbing, i.e., further removed from our ordinary intuitions, than the simple lack of determinism (Majorana in \cite{Bassani:2006}, p.\ 244).
\end{quote}

As one can see, Majorana claims that quantum measurements introduce an ineliminable disturbance so that the initial ontic state of the system cannot be known. That is, such a statement emphasizes that in QM, measurements do not reveal the pre-existing value of the observed property, but rather a certain experimental outcome gives information about the interaction between a given quantum state and an experimental apparatus. Consequently, the above statement may refer to the contextual nature of quantum measurements. 

What is remarkable in this quote, in the opinion of the present author, is the claim that the state before a measurement is not knowable, but not that it is inherently indeterminate, as contemporary supporters of quantum indeterminacy would contend. From this passage, it is plausible to suppose that in Majorana's view QM imposes inherent {\em epistemic} and descriptive limitations, i.e., this theory would restrict our possibility to know and describe the properties of objects existing at microscopic scales. Since the wave function provides the maximal knowledge of physical systems and measurements describe the interaction between such systems and macroscopic devices, one may interpret the ontic state as a substance that exists in the world but cannot be fully known. Thereby, there may be interpretative room to claim that Majorana held a semi-classical conception of quantum objects. 

Two hints seem to go in this direction: not only in the quotation from \cite{Esposito:2006}, p.\ 118 he claimed that $\psi$ provides the probability that the electron {\em is located} in a certain volume of space, but also in his last lecture on QM Majorana affirms that in some cases a quantum object can be intended as a substance with definite properties: 

\begin{quote}
In quantum mechanics, which is dominated by the tendency to reject any objective model, a ``physical quantity'' can strictly speaking be defined only by assigning a method for its measurement. We will admit, however, subject to a posteriori justification, that in some simpler cases, different methods that would classically lead to the same result are indeed equivalent: that is, they lead to the definition of the same physical quantity. In this sense, we will speak of the ``position'' or ``momentum'' of an electron without specifying their methods of observation. And, evidently, it should be interpreted in this sense, Born's statement that $\bar{\psi}\psi$ represents the probability that the electron ``is found'' at a certain point. This observation allows us to easily extend the statistical interpretation of the wave function to deduce the probability that the electron's momentum or its velocity has a given value (\cite{Esposito:2006}, p.\ 119, my translation from the original Italian).
\end{quote}

\noindent However, to propose such an interpretation of Majorana's conception of quantum ontology lies outside the scope of the present text, and it would remain to be demonstrated with full historical rigor, a task that we will leave for another occasion.
 
The second conceptual revolution of QM, as mentioned in the above quotation, concerns the rejection of the deterministic character of classical laws. As Majorana mentions in this paper---and in other places as well, e.g.\ the preparatory notes for his conference analyzed in the previous section---according to classical mechanics, the motion of a physical body is completely determined by the initial conditions of its position and momentum. From the successful application of Newton's law, physicists then tried to generalize the idea according to which every physical body moves obeying universal laws without any exception, in a way that its motion at a certain time is completely determined by the state in which it was at an earlier time. In this way, a mechanistic view of nature was given, in which even the future of a given system is already ``implicit in the present, in the sense that it can be predicted with absolute certainty if the actual state of the universe is entirely known'' (Majorana in \cite{Bassani:2006}, p.\ 239). However, Majorana says determinism irremediably contradicts not only our subjective experience, but also QM, the most advanced theory of physics.\footnote{In this paper, Majorana offers an elegant discussion of the role of statistical laws in classical physics. He also discusses the relationships between the deterministic character of classical mechanical laws and the laws of social sciences.}
In this specific regard, it is relevant to underline that 

\begin{quote}
[t]here are no laws in nature which express a fatal succession of phenomena; even the ultimate laws governing elementary phenomena (atomic systems) are statistical in nature, allowing us to establish only the {\em probability} that a measurement performed on a system prepared in a given way will give a certain result, regardless of the means at our disposal to determine the initial state of the system as accurately as possible. These statistical laws indicate a real lack of determinism and have nothing in common with classical statistical laws in which the uncertainty of results derives from the voluntary renunciation, for practical reasons, of investigating the initial conditions of physical systems in minute detail (Majorana in \cite{Bassani:2006}, p.\ 244, my translation from the original Italian).
 \end{quote}

An example that illustrates the lack of determinism in QM is radioactive decay, which Majorana briefly discussed in his paper. More importantly, he underlines on the one hand, that quantum theory entails the existence of physical phenomena governed by stochastic laws that cannot be reduced or complemented with any causal mechanism. On the other hand, he emphasizes that the introduction of these inherently probabilistic, statistical laws is key to rethinking the relationship between physics and social sciences, which are in turn based on statistical laws as well (cf.\ \cite{Fano:1991}). Rather than thinking about statistical laws in classical terms, as a necessity to model and describe complex systems (about which we often lack a significant amount of data), Majorana speculates about a new reading of the statistical nature of laws present in social sciences. Indeed, he argues, there are no principled scientific arguments that prevent the possibility that human events---as Majorana called them---may be generated by intrinsically stochastic causes, or that they may be genuinely probabilistic in essence. If this hypothesis is considered correct, then the statistical nature of social laws would {\em not} only be due to the necessity to establish empirically generalizations that take into account a great number of variables, but rather such probabilistic character would describe reality {\em as it is}. 

Let us then close this section with Majorana's own words concerning the stochastic nature of laws in social sciences:

\begin{quote}
[f]rom a scientific point of view nothing prevents one from considering that an equally simple, invisible and unpredictable vital fact could be found at the origin of human events. If this is so, as we believe it is, the statistical laws of social sciences increase their function. Their function is not only of empirically establishing the resultant of a great number of unknown causes, but, above all, it is to provide an immediate and concrete evidence of reality (Majorana in \cite{Bassani:2006}, p.\ 258). 
 \end{quote}

\section{Conclusive remarks}
\label{Sec5}
 
After a short biographical note, this paper begins with an overview of Esposito's interpretation of Majorana's perspective on the quantum state. According to this interpretation, which has become widely accepted among historians of science, the preliminary text for his conference in Naples would include the physical foundations of Feynman's path integral formulation of quantum theory. After thoroughly reviewing Esposito's arguments, I examined the textual evidence supporting Majorana's supposed anticipation of Feynman's ideas and argued that Esposito's thesis is not confirmed by such evidence. Alternatively, I contend that these preparatory notes do not anticipate the physical aspects of the path integral approach. Instead, it is shown in detail that Majorana not only referenced Heisenberg's picture of quantum mechanics but also presented it elegantly and concisely while attempting to explain what in classical mechanics would correspond to the notion of quantum state.

Furthermore, by taking into consideration the lecture notes on QM, it has been argued that Majorana held an epistemic interpretation of the wave function. While the quantum state represents an ontic state of affairs, $\psi$ refers to our knowledge of a certain quantum system. Therefore, in his view, the concepts of wave function and quantum state are distinct. Finally, I briefly discussed the major innovations introduced by quantum mechanics from Majorana's perspective, notably the impossibility of extending the objectivity and determinism of classical mechanics to the quantum realm. Majorana states that there are phenomena and laws governing them that are fundamentally statistical and not reducible to causal mechanisms. In this context, the inherent stochastic nature of QM helps to understand the nature of statistical laws in social sciences, since these disciplines also may involve inherent random processes, paving the way for a true unification of natural and social sciences.

Let me conclude by saying that although this chapter did not follow the general attitude that many historians have towards Majorana, i.e.\ the tendency to magnify his pioneering physical intuition, this by no means downplays or diminishes the greatness of Majorana's genius and work. On the contrary, in a book that celebrates the first one hundred years of QM, this chapter honors Majorana's deep understanding, knowledge, and mastery of quantum theory. 
\vspace{5mm}

\noindent {\bf Acknowledgments}: The author is grateful to the Fundação para a Ciência e a Tecnologia (FCT) for financial support (Grant no.\ 2023.07796.CEECIND;  the DOI related to the research project is https://doi.org/10.54499/2023.07796.CEECIND/CP2891/CT0016). Many thanks go to Raoni Arroyo and Matteo Morganti for their kind invitation to contribute to this volume celebrating the birth of quantum mechanics. I would also like to thank Professor Sigfrido Boffi, who gave me great insights into Heisenberg's matrix mechanics, Olga Sarno for her comments on this draft, Gabriele Carcassi and Michele Pizzochero for their useful technical advice, and the audience in Rome for helpful feedback, in particular Vincenzo Fano and Marco Giovannelli. 

%
%
%
%
%

\bibliographystyle{apalike}
\bibliography{PhDthesis}
\end{document}